%% file: paper.tex
\documentclass[journal,twoside,web]{color}
\usepackage{generic}
\usepackage{cite}
\usepackage{amsmath,amssymb,amsfonts}
\usepackage{bbold}
\usepackage{algorithmic}
\usepackage{graphicx}
\usepackage{textcomp}
\usepackage{soul, color}
\usepackage{steinmetz}
\usepackage{comment}
{\begin{list}{$\bullet$ \hfill}{
			\setlength{\leftmargin}{\parindent}
			\setlength{\parsep}{0.07\baselineskip}
			\setlength{\itemsep}{0.7\parsep}
			\setlength{\labelwidth}{\leftmargin}
			\setlength{\labelsep}{0em}}}{\end{list}}
\def\BibTeX{{\rm B\kern-.05em{\sc i\kern-.025em b}\kern-.08em
    T\kern-.1667em\lower.7ex\hbox{E}\kern-.125emX}}
{Soylu and Oelze: Calibrating Data Mismatches in Deep Learning for QUS by Using \textit{Setting Transfer Functions}}
\begin{document}
\title{Calibrating Data Mismatches in Deep Learning-Based Quantitative Ultrasound Using \textit{Setting Transfer Functions}}
%Velocity Filtering for Improved Resolution \\ in Ultrasound Localization Microscopy}
\author{Ufuk Soylu and Michael L. Oelze
\thanks{This work was funded by grants from the NIH (R01CA251939, R21EB023403 and R21EB024133).}
\thanks{Ufuk Soylu and Michael Oelze are with the Department of Electrical and Computer Engineering, and the Beckman Institute, University of Illinois at Urbana-Champaign, Urbana, IL 61801 USA. 
(e-mail: usoylu2@illinois.edu; oelze@illinois.edu).}
}
\maketitle
\begin{abstract}
\input{abs}

\end{abstract}

\input{intro}
\input{theory}

\input{methods}
\input{results}
\input{discussions}

\input{conc}
% \section*{Acknowledgment}
\bibliographystyle{IEEE_ECE}
% Put references in BibTeX format in thesisrefs.bib.
\bibliography{paper}
%\todo[inline]{The following applies to about 10 of the references: "IEEE publications must list names of all authors, up to \emph{six} names. If there are more than six names listed, use the primary author’s name followed by et al." (See https://ieeeauthorcenter.ieee.org/wp-content/uploads/IEEE-Reference-Guide.pdf) --Done.}
%\todo[inline]{Use the standard abbreviations for publication names, e.g., 
%"IEEE Transactions on Ultrasonics, Ferroelectrics, and Frequency Control" =
%"IEEE Trans. Ultras. Ferroel. Freq. Cont.", etc. You can get an extra bib file with all the standard abbreviations and include it. For some sources, you can use extreme abbreviations (edit your bib file manually), because we are submitting to an IEEE journal and all readers would know what "ICASSP 2017" is:
%"2017 IEEE International  Conference  on  Acoustics,  Speech  and  Signal  Processing(ICASSP).    IEEE, 2017" =
%"ICASSP 2017"--Done}
% \include{supp}
\end{document}

%% file: abs.tex
Deep learning (DL) can fail when there are data mismatches between training and testing data distributions. Due to its operator-dependent nature, acquisition-related data mismatches, caused by different scanner settings, can occur in ultrasound imaging. Therefore, mitigating effects of acquisition-related data mismatches is essential for wider clinical adoption of DL powered ultrasound imaging and tissue characterization. To mitigate the effects, ideally we need to collect a large training set at each scanner setting. However, acquiring such training sets is expensive. Another approach could be training on a subset of imaging settings, which makes the data generation less expensive. However, there will still be generalization issues for the settings that are not included in the training. As an alternative approach that is inexpensive and generalizable, we propose to collect a large training set at a single setting and a small calibration set at each scanner setting. Then, the calibration set will be used to calibrate data mismatches by using a signals and systems perspective. We tested the proposed solution to classify two phantoms using an L9-4 array connected to a SonixOne scanner. To investigate generalizability of the proposed solution, we calibrated three types of data mismatches: pulse frequency mismatch, focus mismatch and output power mismatch. For the training and the testing sets, a video of 1,007 frames per phantom was captured by free-hand motion. For the calibration set, 10 frames per phantom were captured after stabilizing the transducer. A well-known convolutional neural network (CNN), i.e., ResNet-50, was trained using the ultrasound RF data. To calibrate the setting mismatches, we calculated the \textit{setting transfer functions}. The CNN trained with no calibration resulted in mean classification accuracies of 55.3\%, 64.4\% and 70.3\% for pulse frequency, focus and output power mismatches, respectively. By using the \textit{ setting transfer functions}, which allowed a matching of the training and testing domains, we obtained mean accuracies of 95.3\%, 92.99\% and 99.32\%, respectively. Therefore, the incorporation of the \textit{setting transfer functions} between scanner settings can provide an economical means of generalizing a DL model for specific classification tasks where scanner settings are not fixed by the operator. 
%We obtained transfer functions in two ways: (i) from a phantom which is not included in the training, dubbed as “label-free”, (ii) from the classification phantoms, dubbed as “label-dependent”. 
%To create a pulse frequency mismatch, we used 9 MHz and 5 MHz pulse frequency settings in the training and in the test, respectively. To create a focus mismatch, we used training data focused axially at 2 cm  and test data focused axially at 1 cm and 3 cm. To create an output power mismatch mismatch, we used 0 db and -6 db output power settings in the training and in the test, respectively.  

%% file: intro.tex
\section{Introduction}
\label{sec:intro}

Deep learning (DL) powered biomedical ultrasound imaging is becoming more advanced and coming closer to routine clinical applications in recent years due to a rapid increase in computational power and greater availability of large datasets \cite{akkus2019survey}. DL is the process of learning a hierarchy of parameterized nonlinear transformations to perform a desired function, and therefore, DL extracts a hierarchy of features from raw input images automatically rather than extracting features manually. Among DL algorithms, convolutional neural networks (CNN) are the most popular structure for ultrasound biomedical imaging because they are data-efficient learners for image analysis tasks thanks to their translational invariance \cite{liu2019deep}.

One application of DL in diagnostic ultrasound is the classification of tissue state from raw radiofrequency (RF) ultrasound backscatter. In an initial work, a CNN was used to classify liver steatosis in rabbits based on the RF backscattered data \cite{nguyen2019reference}. The advantage of the approach was that the CNN learned the tissue signal and separated it from the system signal. Therefore, the CNN approach did not require a separate calibration spectrum for analysis and did not need to fit the data to a pre-conceived model, i.e., the CNN learned the model. This CNN approach was then compared to a traditional quantitative ultrasound approach in the characterization of fatty liver with better performance attributed to the CNN \cite{nguyen2021use}. Similar approaches were later used to quantify liver disease in human patients and to characterize breast masses for cancer detection \cite{han2020noninvasive, byra2022liver,byra2022joint}. 

Even though DL is promising for classification of tissues based on their RF backscattered data; there are two main roadblocks to wider clinical adoption as Castro et al stated in \cite{castro2020causality}. One road block to deploying DL powered algorithms to real clinical settings is data scarcity. Specifically, there is scarcity of labelled data, in great part due to high costs of conducting lab experiments or acquiring expert annotations. The other main road block is data mismatch. Data mismatch happens due to mismatches between development and deployment environments and tends to limit generalizability of DL-based algorithms. Overall, in the generic case, when there is not enough labelled data or when no assumptions can be made about the mismatches between development and deployment settings, any learning-based algorithm would be ineffective. Therefore, to turn DL powered biomedical ultrasound imaging into reality, there is a constant need for developing DL algorithms, which are data efficient and more robust against data mismatches in ultrasound images.  

The clinical environment is too complicated to be fully realized in a development setting where operators, such as sonographers, are adjusting settings to obtain the best perceived image quality. Therefore, there will be inevitable data mismatches if the operator is given the freedom to select settings that provide perceived optimal image quality and these settings do not match scanner settings associated with the training data. Specifically, we consider acquisition-related data mismatches, i.e., data mismatches caused by variations in scanner  settings such as the number of foci and their locations or pulse frequency, and develop a method by looking at the problem from a signals and systems perspective. Such data mismatches are pervasive in biomedical ultrasound imaging due to its operator-dependent and patient-dependent nature and mitigating their effects is essential for wider clinical adoption of DL-based methods  for tasks such as tissue classification.

In this work, we propose a method to mitigate the effects of data mismatch in biomedical ultrasound imaging with the specific task of characterizing tissues based on  raw RF data derived from ultrasonic backscatter. The proposed method is an inexpensive way of mitigating generazibility issues caused by acquisition-related data mismatches in biomedical ultrasound imaging. In our experiments, we chose tissue classification or quantitative ultrasound (QUS) as the primary application and tested our proposed method to classify distinct tissue-mimicking phantoms. Traditional QUS approaches can account for system- and operator-dependent changes to the settings through calibration of the system at each setting \cite{nam2012cross, wirtzfeld2010cross}. QUS has recently evolved from model-based approaches \cite{kemmerer2012ultrasonic, oelze200711b} to model-free, DL-based techniques \cite{nguyen2019characterizing,nguyen2019reference,nguyen2021use,soylu2022data}. However, these DL-based techniques used a single scanner setting. Similar to those studies, we also use DL algorithms with the backscatterd RF ultrasound signal to utilize the phase and frequency-dependent information for the tissue classification problem without the constraint that the settings be the same for each scan. The theoretical bases for this approach are discussed in \ref{sec:theory}. Further details of our experiments can be found in Section \ref{sec:met} and \ref{sec:results}. We provide discussions and conclusions in Section \ref{sec:discussion} and \ref{sec:conclusion}, respectively.

%% file: theory.tex
\section{Theory}
\label{sec:theory}
Any form of learning algorithm would be ineffective if there are data mismatches between training and testing data, and no assumptions can be made to calibrate the mismatches. Ideally, we need to collect a large and diverse training data set at each imaging setting to completely eliminate data mismatches caused by scanner parameters. However, acquiring such a training data set can be extremely expensive. Another approach could be training on a subset of imaging settings, which makes the data generation and gathering process less expensive. However, there will still be generalization issues for the settings that are not included in the training set. The question we address here is "How can we mitigate acquisition-related data mismatches in an inexpensive and generalizable way for QUS?"

To accomplish this, we consider a systems response approach. Biomedical ultrasound imaging can be viewed as a system, which encodes all the information related to an imaging system working on a tissue signal, which encodes all the information related to an imaging substrate. Subsequently, an image obtained by an ultrasound imaging system can be decomposed into two parts: a system response and a tissue signal. Then, the problem of acquisition-related data mismatch can be posed as matching system responses. When training inputs and testing inputs are from different distributions due to different scanner settings, such as a different excitation pulse frequency, there are two different system responses in the imaging process, one corresponding to training and the other one corresponding to deployment or testing. Potentially, a function can be defined that allows a system to transfer from one environment to the other environment. 

More precisely, under a single scattering approximation and when at least one aperture diameter away from the transducer surface, the backscattered frequency spectrum from a medium can be represented as \cite{mamou2013quantitative}
\begin{align}
\label{powerspec_initial}
    W(f,\textbf{x}) = T(f,\textbf{x}) A(f,\textbf{x}) D(f,\textbf{x}) H(f) R(f,\textbf{x})
\end{align}
where $f$ represents frequency, $x$ represents axial direction, $T (f,\textbf{x})$ incorporates the transmission losses between tissues, $A(f,\textbf{x})$ is the frequency-dependent attenuation, $D(f,\textbf{x})$ represents the diffraction effects of the transducer, $H (f)$ is the impulse response of the transducer system and incorporates the electro-mechanical response, and $R(f,\textbf{x})$ is the scattering function describing the underlying tissue micro-structure. Therefore, an ultrasound image can be naively decomposed as 
\begin{align}
\label{powerspec_simple}
    W(f,\textbf{x}) = S_\phi(f,\textbf{x}) P(f,\textbf{x})
\end{align}
where $S_\phi(f,\textbf{x}) = D(f,\textbf{x}) H(f,\textbf{x}) $ is the system response which incorporates all the information related to ultrasound imaging system and $P(f,\textbf{x}) = T(f,\textbf{x}) A(f,\textbf{x})R(f,\textbf{x})$, which is the tissue or sample signal that incorporates all the information related to imaging substrate (i.e., attenuation, transmission losses and scattering function). The subscript $\phi$ in $S_\phi(f,\textbf{x})$ represents the scanner setting, exclusively $\phi_{train}$ stands for training environment and $\phi_{test}$ stands for testing environment, for later use. To calibrate the acquisition-related data mismatches between two system settings, we setup the scenario where the tissue signal, $P(f,\textbf{x})$, does not change between testing and training, such that,
\begin{align}
\label{powerspec}
   \frac{W_{test}(f,\textbf{x})}{W_{train}(f,\textbf{x})} &= \frac{S_{\phi_{test}}(f,\textbf{x})}{S_{\phi_{train}}(f,\textbf{x})}\\ &= \Gamma(f,\textbf{x})
\end{align}
where $\Gamma(f,\textbf{x})$ represents the\textit{ "setting transfer function" }between training and testing system parameters. This can be done in a practical way by selecting a tissue mimicking phantom with uniform scattering properties and fixing the transducer to scan and record the signal from a single location in the phantom while the settings are changed from training to testing. To calibrate in training time,  $\Gamma(f,\textbf{x})$ is sufficient:
\begin{align}
\label{training time}
    W_{test}(f,\textbf{x}) &= \Gamma(f,\textbf{x}) W_{train}(f,\textbf{x})
\end{align}
and hence, it is a convolution operation in time direction, $t$ 
\begin{align}
\label{training time_spatial_domain}
    w_{test}(t,\textbf{x}) &= \gamma_{train}(t,\textbf{x}) \ast_{t} w_{train}(t,\textbf{x})
\end{align}
where $\gamma_{train}$ represents the \textit{"setting transfer filter"} for training time. For the train-time calibration, a DL model is trained at testing data settings, and therefore, testing data can be input into the DL model directly. In order to achieve this, the training data are converted to the testing data via the setting transfer function $\Gamma(f,\textbf{x})$ in training time and a new model is developed by the network, which allows the testing data settings to match with training settings used to create the model. 

Similarly, $\Gamma^{-1}(f,\textbf{x})$ is sufficient for calibrating in testing time:
\begin{align}
\label{test time}
   W_{train}(f,\textbf{x}) &= \Gamma^{-1}(f,\textbf{x}) W_{test}(f,\textbf{x}),
\end{align}
which results in the following filtering operation
\begin{align}
\label{test time_spatial_domain}
    w_{train}(t,\textbf{x}) &= \gamma_{test}(t,\textbf{x}) \ast_{t} w_{test}(t,\textbf{x})
\end{align}
where $\gamma_{test}$ represents the \textit{"setting transfer filter"} for testing time. For the test-time calibration, a DL model is trained at the training data settings, and therefore, testing data can't be input into the DL model directly. The testing data needs to be converted to training data via the setting transfer function $\Gamma^{-1}(f,\textbf{x})$. Then, the model developed with the original training data and its associated settings are used with the converted testing data.  

In this work, we propose to use a signals and systems perspective to \textit{calibrate} training or testing data. Here is the proposed method step by step at a high-level:

\begin{itemize}
    \item[1.] Generate and gather a \textit{large} amount of training data at a single setting.
    \item[2.] Generate and gather a \textit{small} amount of calibration data at each scanner setting.
    \item[3.] Calculate $W_{test}$ or $W_{train}$ by using the calibration set to calculate the setting transfer function $\Gamma$ and $\Gamma^{-1}$
    \item[4.] Construct linear phase filters $\gamma_{train}$ or  $\gamma_{test}$ by using the magnitude responses of $\Gamma$ and $\Gamma^{-1}$.
    \item[5.] When a different setting than training setting is being used in the scanning process, calibrate the mismatch by using filters $\gamma_{train}$ or $\gamma_{test}$ either in the training time or in the testing time in the DL network, respectively.  
\end{itemize}

%% file: methods.tex
\section{Methods}
\label{sec:met}
\subsection{Phantoms}
\begin{figure}[hbt!]
\begingroup
    \centering
    \begin{tabular}{c c c}
\hspace{-0.2cm}{ \includegraphics[scale=0.041]{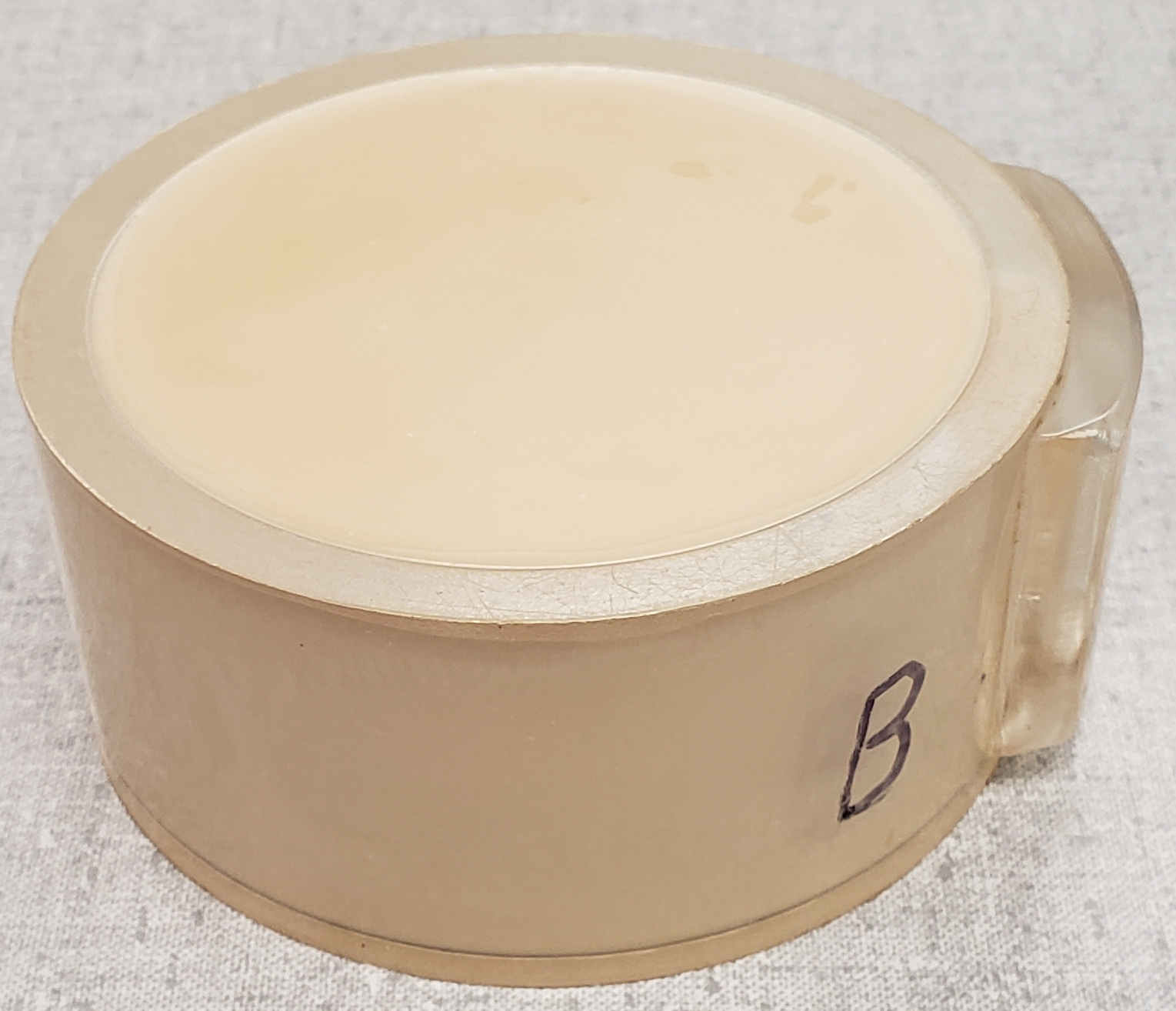}}&\hspace{0cm}{ \includegraphics[scale=0.036]{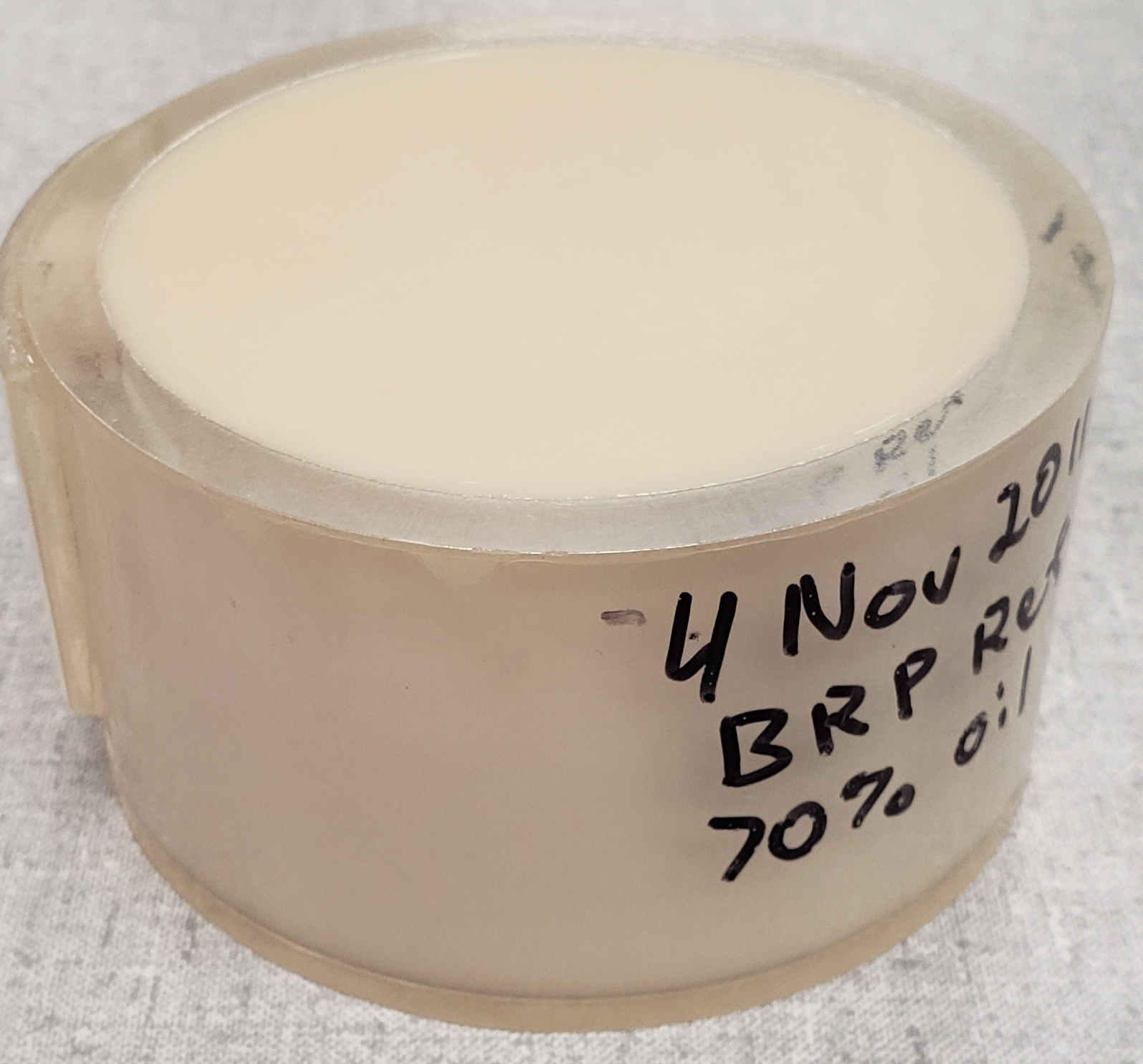}}& \hspace{-0.1cm}{ \includegraphics[scale=0.031]{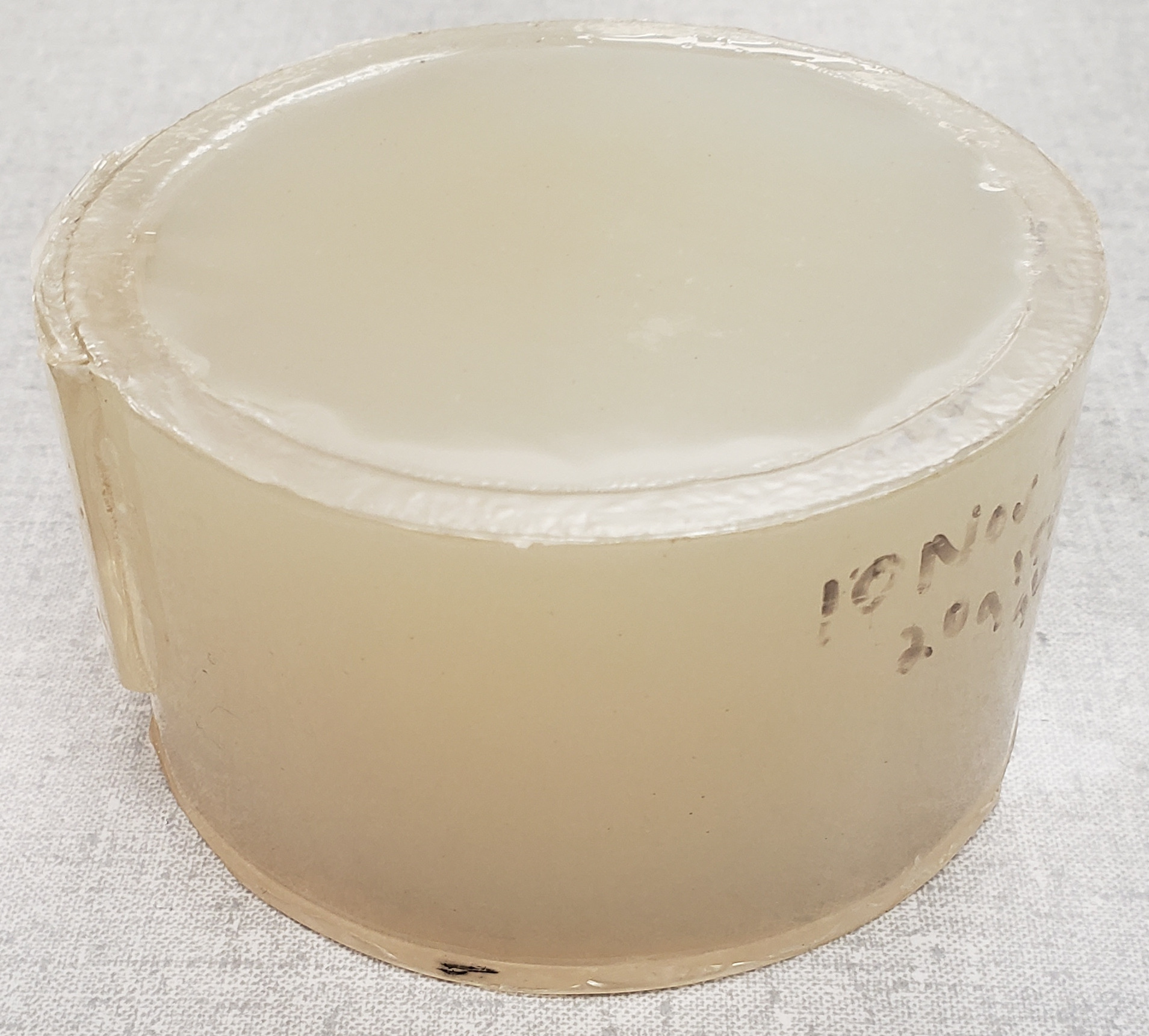}} \\
\hspace{-0.5cm} Phantom1 &\hspace{-0cm} Phantom2 &\hspace{-0.1cm}Calibration\\
\end{tabular}
\caption{Photographs of The Classification and Calibration Phantoms}
\label{fig:phantom_picutes}
\endgroup
\end{figure}

Two different tissue-mimicking phantoms were classified in the experiments, which we designated as Phantom1 and Phantom2, and another phantom was used as the calibration phantom. They are cylindrically shaped as shown in Fig. \ref{fig:phantom_picutes}.

Phantom1, which mimics soft tissue, has been described by Wear et al. \cite{wear2005interlaboratory}. 
Their materials were produced based on the method of Madsen et al. \cite{madsen1998liquid} and they are macroscopically uniform. The only nonuniformity results from the random positioning of microscopic glass bead scatterers. Phantom1 had a measured attenuation coefficient slope of approximately 0.7 dB$\times$cm$^{-1}\times$MHz $^{-1}$, respectively. The component materials and their relative amounts by weight for Phantom1 are agarose (2.34$\%$), n-propanol (2.92$\%$), 75 to 90 $\mu$m-diameter glass beads (1.87$\%$), bovine milk concentrated 3 times by reverse osmosis (47.9$\%$), liquid Germall Plus preservative (International Specialty Products, Wayne, NJ) (1.87$\%$), and 18-$M\Omega$-cm deionized water (43.1$\%$).

Phantom2 has been described by Nam et al. \cite{nam2012cross} as a reference phantom. Its measured attenuation coefficients at frequencies from 2 to 10 MHz were fit to a power law function of frequency, $\alpha(f) = 0.256 f ^{1.366}$, where $f$ is the frequency in terms of MHz and $\alpha(f)$ is in terms of dB/cm. The phantom was made with 6.4 g of 5-43 $\mu$m-diameter glass beads uniformly distributed spatially at random in a gel background. The background material was a gelatin emulsion containing 70$\%$ safflower oil \cite{madsen2006stability}.

The calibration phantom was a low attenuation phantom, which was constructed as described by Anderson et al. \cite{anderson2010interlaboratory}. It had a weakly scattering 2\% agar background with 150-180 $\mu$m glass beads, which had a slightly broader distribution of scatterer sizes (160 $\pm$ 60 $\mu$m). The glass bead concentration was 20 g/L and the beads were randomly distributed spatially within the phantom.

\vspace{-0.3cm}
\subsection{Ultrasound Imaging Device and Imaging Settings}
Ultrasound gel was placed on the surfaces of the phantoms and then the phantoms were scanned with an L9-4/38 transducer using a SonixOne system (Analogical Corporation, Boston, MA, USA) providing an analysis bandwidth of 2-7.5 MHz and focusing with an f-number of 3 for transmit and receive. Ultrasound frames of post-beamformed RF data sampled at 40 MHz were acquired from each of the phantoms and saved for offline processing. The ultrasound post-beamformed RF data were directly used in the training, test and calibration processes. The imaging array had a center frequency measured at 5.5 MHz and was operated with a fixed elevational focus of 1.9 cm. Scanner parameters that were adjusted for each experiment can be found in Table \ref{tab:imaging_settings}. Changes in the focus occurred on transmit. We acquired data from the phantoms via two scanning procedures. In the first, we recorded a video of 1,007 ultrasound frames by free-hand motion. Free-hand acquisition provided us a large data set of independent frames for each phantom to be used in the training and testing. In the second procedure, we stabilized the transducer using a bar clamp holder and then recorded 10 frames at both the training and testing settings from the exact same location in the phantom which provided us with the calibration data to be used in calculating the setting transfer functions.    

\begin{table}[htb!]
    \centering
\caption{Scanner Parameters for SonixOne System}
{\renewcommand{\arraystretch}{1.1}
{\footnotesize
\begin{tabular}{ |p{2.75cm}||p{1cm}||p{1.2cm}||p{1cm}|}
 \hline
& Pulse Freq& Focus&Output Power \\
 \hline
Training in Sec.\ref{sec:results_pulse}& 9 MHz &@2cm&0 dB\\
Test in Sec.\ref{sec:results_pulse}& 5 MHz &@2cm&0 dB\\
\hline
Training in Sec.\ref{sec:results_focus}& 9 MHz&@2cm&0 dB \\
Test in Sec.\ref{sec:results_focus}&9 MHz &@1cm and 3cm&0 dB\\
\hline
Training in Sec.\ref{sec:results_mix}& 9 MHz& @2cm&0 dB\\
Test in Sec.\ref{sec:results_mix}& 9 MHz& @2cm&-6 dB\\
 \hline
\end{tabular}
}}
\label{tab:imaging_settings}
\end{table}
\vspace{-0.3cm}
\subsection{Data-Set}
\label{dataset}

The total size of an ultrasound image frame from the phantoms was 2,080 pixels $\times$ 256 pixels. There were 2,080 samples along the axial direction that corresponded to a 4 cm imaging depth. Even though the L9-4/38 transducer has 128 channels, the SonixOne system interpolates to 256 channels that correspond to 256 lateral samples. After acquiring ultrasound frames by either stable acquisition or free-hand acquisition, we extracted square data patches from the image frames whose sizes were 256 samples $\times$ 256 samples, to be used in training, validation and testing sets. The data used in the DL network were the raw backscattered RF data. The data-set of ultrasound frames is also publicly available at https://figshare.com/s/7ae94a537a56e5db3525.

\begin{figure}[hbt!]
\begingroup
    \centering
    \begin{tabular}{c}
\hspace{-0.0cm}{ \includegraphics[width=0.48\linewidth]{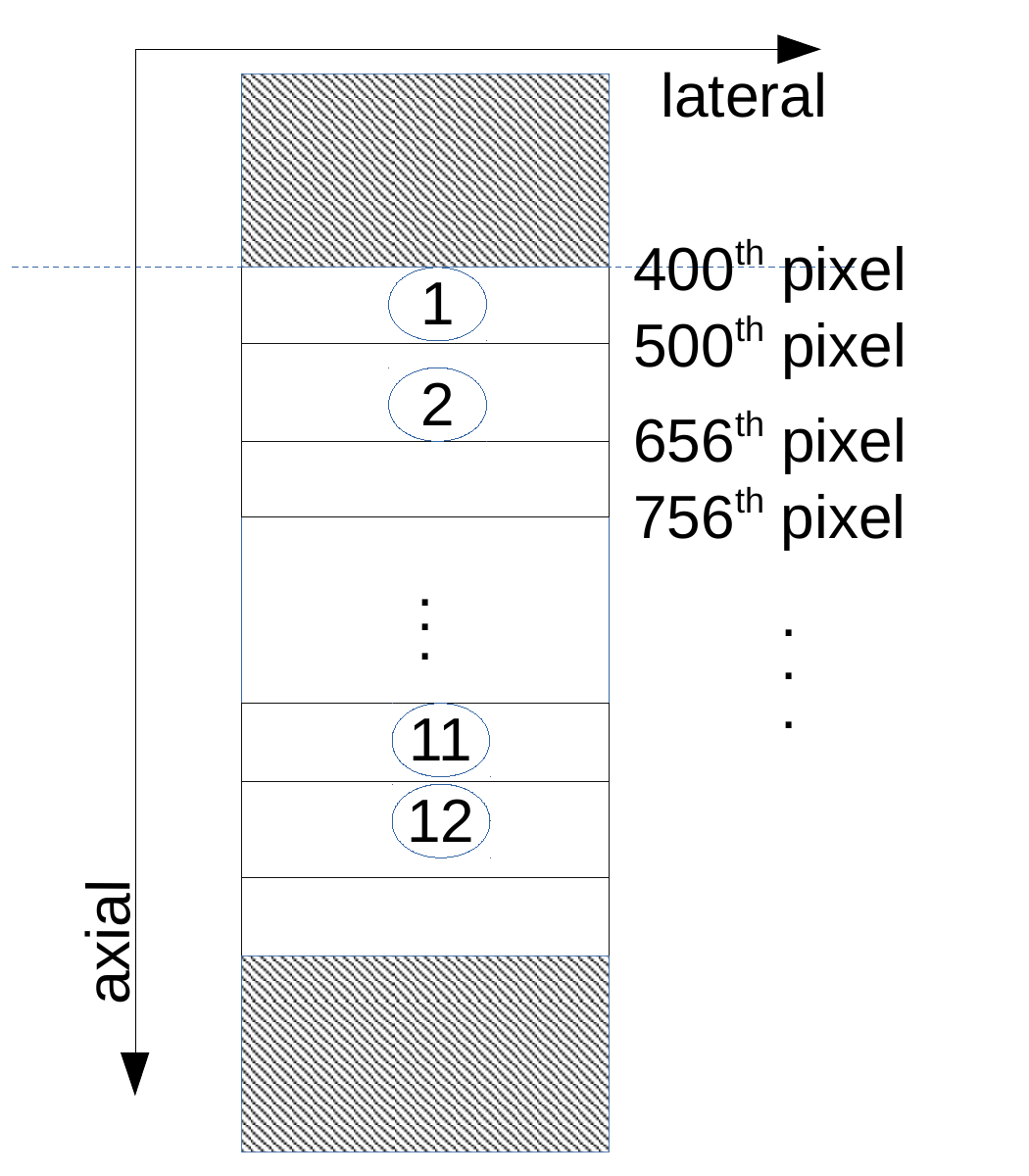}}\\
\end{tabular}
\caption{Patch Extraction: Local patches, whose sizes were 256 $\times$ 256 samples, were extracted to be input into a CNN. The first 400 samples were not used. By incrementing the start of each patch 100 samples axially, 12 patches were extracted per frame.}
\label{fig:patch_extract}
\endgroup
\end{figure}

From one ultrasound image, we extracted 12 image patches as depicted in Fig. \ref{fig:patch_extract}. While extracting image patches, we didn't use the first 400 samples in the ultrasound image. The first patch started at sample 400 axially, the second patch started at sample 500, each subsequent patch starting point was incremented by 100 samples. This corresponds to a 80$\%$ overlap axially between data patches. As a consequence, the training set consisted of square patches extracted from ultrasound frames acquired at scanner settings for the training. On the other hand, ultrasound frames acquired for the test data were split into two sets: one was for the validation set and the other was for the test set. 

In total, by considering two phantoms, we extracted 24,168 patches at scanner settings for the training, which is equal to 1,007 frames per phantom$\times$2 phantoms$\times$12 patches per frame. Then, we randomly selected four fifths of the total patches to be used in the training, which resulted in 19,334 patches. Similarly, after acquiring 2,014 frames at scanner settings for the testing, as the validation and test sets, we randomly selected 750 ultrasound frames out of 2,014 ultrasound frames, which resulted in 9,000 patches for validation and testing. We repeated the random selection of training and testing patches ten times for each experiment.
\begin{comment}
 Additionally, the data-set of ultrasound images is also publicly available at https://osf.io/7ztg3/ (DOI 10.17605/OSF.IO/7ZTG3).
\end{comment}

\vspace{-0.3cm}
\subsection{Network Structure}

In this work, we used a well-known CNN architecture, i.e., ResNet-50 \cite{he2016deep}. CNNs have several advantages among other DL structures for the tasks related to 2D images. CNNs are similar to the human visual system, which makes them effective at learning and extracting abstractions of 2D images \cite{alom2018history}. The ResNet-50 architecture was slightly modified by decreasing the number of input channels to 1 from 3 and the number of output logits to 2 from 1000. 
The architecture is described in Table \ref{tab:network}.

\begin{table}[htb!]
    \centering
\caption{DL Network Architecture}
{\renewcommand{\arraystretch}{1.05}
{\footnotesize
\begin{tabular}{c|c|c}
\hline stage & output & ResNet-50 \\
\hline conv1 & $128 \times 128$ & $7 \times 7,64$, stride 2 \\
\hline & & $3 \times 3$ max pool, stride 2 \\
\cline{3-3} conv2 & $64 \times 64$ & 
{$\left[\begin{array}{l}1 \times 1,64 \\
3 \times 3,64 \\
1 \times 1,256\end{array}\right] \times 3$} \\
\hline conv3 & $32 \times 32$ & {$\left[\begin{array}{l}1 \times 1,128 \\
3 \times 3,128 \\
1 \times 1,512\end{array}\right] \times 4$} \\
\hline conv4 & $16 \times 16$ & {$\left[\begin{array}{l}1 \times 1,256 \\
3 \times 3,256 \\
1 \times 1,1024\end{array}\right] \times 6$} \\
\hline conv5 & $8 \times 8$ & {$\left[\begin{array}{l}1 \times 1,512 \\
3 \times 3,512 \\
1 \times 1,2048\end{array}\right] \times 3$} \\
\hline & $1 \times 1$ & global average pool \\
& & 2-d fc, softmax \\
\hline
\end{tabular}
}}
\label{tab:network}
\end{table}

% In the experiments, we use pre-trained ResNet-18, which was pre-trained on ImageNet dataset as PyTorch \cite{paszke2019pytorch} default, as the initialization point. We then modify the first convolutional layer which takes 3 input feature channels, to work with single feature channel at its input. Additionally, we modified the last layer which is a linear layer. Instead of outputting 1000 classes, the network now outputs 3 classes corresponding 3 phantoms.  
\vspace{0.1cm}
\subsection{Training}

DL training was done separately by using three machines, each with a single GPU. One machine had a TITAN RTX and the other two machines had a RTX A500. All implementations were done with the PyTorch library \cite{paszke2019pytorch}. 

As a data preprocessing step, we applied a simple denoising filter, which filters out the frequencies except the range between 1-10 MHz. For designing the denoising filter, we used Python's scipy.signal.firwin2 function. The frequency sampling points were between 0 MHz to 20 MHz at 0.15625 MHz increments. Filter gain was selected as 1 for the frequencies between 1-10 MHz and 0 for the other frequencies. The filter was implemented via Python's scipy.signal.convolve, whose mode parameter was selected as "same". Then, the models were trained by using cross entropy loss with uniform class weights. Horizontal flip with 0.5 probability was implemented as a default data augmentation step in the training process. The batch number was chosen as 64 for all experiments. We used the Adam algorithm \cite{kingma2014adam} as the optimizer in all experiments. The learning rates and the epoch numbers were adjusted by using the validation set. 

After adjusting all the training parameters, we repeated training and testing for each experiment 10 times starting from random selection of ultrasound frames and dividing them into patches. Then, we calculated the mean classification accuracies and standard deviations in the test set. In Section \ref{sec:results}, we report those numbers for each experiment.  
\vspace{-0.3cm}
\subsection{Calibration}

For our method, a \textit{small} calibration data set obtained at each scanner setting was necessary to account for the data mismatches between the training and the testing environment. We gathered the calibration by stabilizing the transducer array on top of a phantom and acquiring scans from the exact same view for each setting. For the training and testing settings, we acquired 10 frames of the same view from the phantom. Ten frames were used to reduce any systematic noise, such as electrical noise, by averaging.  

By using the calibration set, we obtained the setting transfer function that can transfer an imaging system behavior with a given set of scanner parameters to another set of scanner parameters. To calibrate data mismatches between training and testing environments, we first calculated $W_{train}(f,\textbf{x})$ and $W_{test}(f,\textbf{x})$  in \eqref{powerspec}. Because of the stable acquisition setup, i.e. using the same tissue signal $P(f,\textbf{x})$ in the calibration set, by taking their ratios, we obtain setting transfer functions $\Gamma(f,\textbf{x})$ and $\Gamma^{-1}(f,\textbf{x})$ between training and testing sets. The ratios can be used either in training time by applying $\Gamma(f,\textbf{x})$ as shown in \eqref{training time}, which we call \textit{"train-time calibration"} or in test time by applying $\Gamma^{-1}(f,\textbf{x})$ as shown in \eqref{test time}, which we call \textit{"test-time calibration"}. 

%In short, while the train-time calibration calibrates the model to match with testing data settings, the test-time calibration calibrates the testing data to match with the model. 

%Furthermore, we obtained the setting transfer functions in two ways: (i) from a phantom which is not included in the training, dubbed as \textit{“label-free calibration”}, (ii) from the phantoms used in classification, dubbed as \textit{“label-dependent calibration”}. In \textit{“label-dependent calibration”}, we used phantom specific  setting transfer functions. For instance, we applied the ratio of the transfer functions, which are obtained from Phantom1, to Phantom1. Therefore, \textit{“label-dependent calibration”} requires the label information which is only achievable in training time.

In practice, we implemented the setting transfer functions in a manner inspired by the Wiener filter,
\begin{align}
\label{wiener train}
    \Gamma_{Wiener}(f,\textbf{x}) &= \frac{\frac{1}{|\Gamma(f,\textbf{x})|}}{\frac{1}{|\Gamma(f,\textbf{x})|^2}+\frac{1}{SNR}}
\end{align}

\begin{align}
\label{wiener test}
    \Gamma^{-1}_{Wiener}(f,\textbf{x}) &= \frac{|\Gamma(f,\textbf{x})|}{|\Gamma(f,\textbf{x})|^2+\frac{1}{SNR}}
\end{align}
where $SNR$ was estimated through the power spectra of $W_{train}$ and $W_{test}$. First, we determined a noise floor level by looking at the lowest values of the power spectra outside of the transducer bandwidth, i.e., around 20 MHz. Then, at each frequency bin, we calculated the tissue signal level by subtracting the noise floor. Subsequently, we obtained SNR values by taking the ratios of the tissue signal and the noise floor at each frequency bin for both the power spectra of $W_{train}$ and $W_{test}$. Lastly, we took the minimum SNR value between $W_{train}$ and $W_{test}$ used that value in the filter. In this filter design, at frequency bins when the signal level of the setting transfer function was small compared to the noise, the filter acted like a denoising filter. When the signal level of the setting transfer function was high compared to noise, the filter was equivalent to the original $\Gamma$. This provides a robust way to use the complete bandwidth of setting transfer functions.

After patch extraction, in the calibration process, we first obtained square patches for training and testing, whose sizes were 256 axial samples $\times$ 256 lateral samples, coming from different axial locations for each ultrasound frame in the calibration set. At this point, for each scanner setting and depth location, we obtained 10 square patches because we acquired 10 frames by stable acquisition for each scanner setting in the calibration set. Next, we calculated a one dimensional power spectrum along the axial direction for each scan line (channel) of data and collected 256 power spectra per square patch. Then, we took the average of the power spectra channel-wise and frame-wise to come up with a single power spectrum representing the imaging process with the given setting at each axial location. By doing so, we obtained depth-dependent power spectra at training and testing conditions. In mathematical terms, we obtained $\{|W_{train}(f)|^2_{1}, |W_{train}(f,\textbf{x})|^2_{2},\hdots, |W_{train}(f,\textbf{x})|^2_{12} \}$ and $\{|W_{test}(f,\textbf{x})|^2_{1}, |W_{test}(f,\textbf{x})|^2_{2}, \hdots, |W_{test}(f,\textbf{x})|^2_{12}\}$. The subscripts here denote the axial patch number. 

The power spectra, and hence the setting transfer functions, don't depend on the lateral direction because the scattering properties of the phantom were uniform. Therefore, we took the average of the power spectra channel-wise, i.e., laterally. Next, we obtained the setting transfer functions $\Gamma(f,\textbf{x})$ or $\Gamma^{-1}(f,\textbf{x})$ by taking the square root of the ratio of the power spectra between the training and testing settings. Subsequently, we obtained $\Gamma_{Wiener}(f,\textbf{x})$ or $\Gamma^{-1}_{Wiener}(f,\textbf{x})$. After calculating $\Gamma_{Wiener}(f,\textbf{x})$ or $\Gamma^{-1}_{Wiener}(f,\textbf{x})$, we used scipy.signal.firwin2 function from Python, which constructs an FIR filter with linear phase from the given frequencies and corresponding gains, to obtain the setting transfer filters $\gamma_{train}$ from \eqref{training time_spatial_domain} or $\gamma_{test}$ from \eqref{test time_spatial_domain}. The number of taps in the FIR filter was selected as 151. The frequency sampling points were between 0 MHz to 20 MHz at 0.15625 MHz increments. The desired filter gains of $\Gamma_{Wiener}(f,\textbf{x})$ or $\Gamma^{-1}_{Wiener}(f,\textbf{x})$ at the frequency sampling points were inputs to the Python function. Then, the implementation of convolution operations in \eqref{training time_spatial_domain} and \eqref{test time_spatial_domain} were done via the Python function torch.nn.functional.conv1d whose parameter padding was selected as 'same', which pads the input so the output has the same shape as the input.

\begin{figure}[hbt!]
\begingroup
    \centering
    \begin{tabular}{c}
\hspace{-0.4cm}{ \includegraphics[width=8.8cm, height =4.2cm]{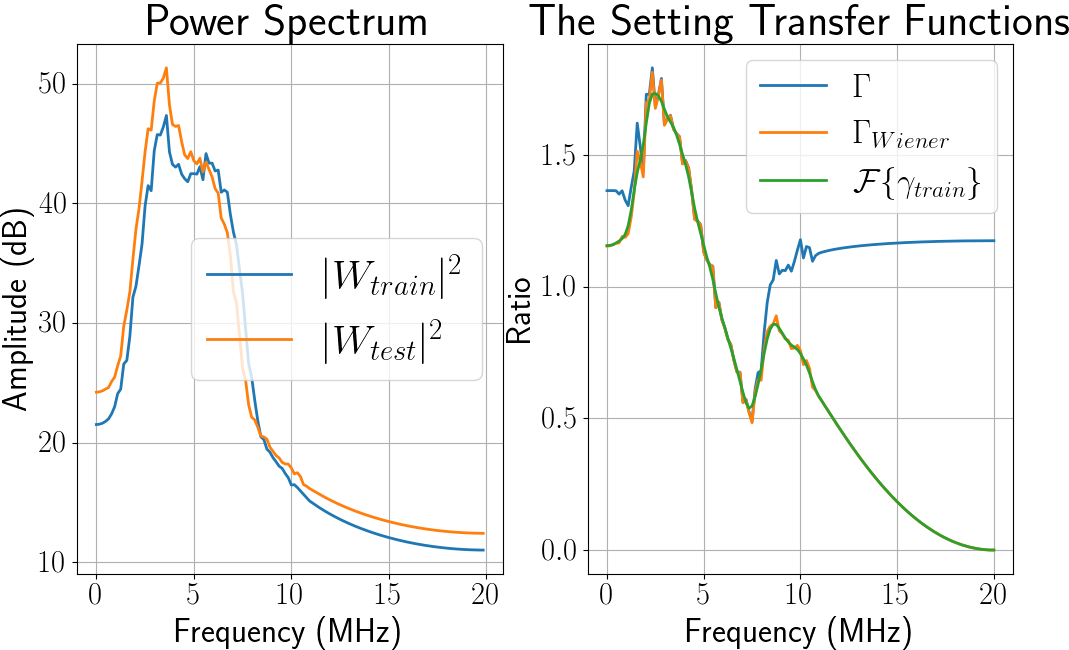}}\\
\end{tabular}
\caption{Calibration plots for the pulse frequency mismatches}
\label{fig:calib1}
\endgroup
\end{figure}

\begin{figure}[hbt!]
\begingroup
    \centering
    \begin{tabular}{c}
\hspace{-0.4cm}{\includegraphics[width=8.8cm, height =4.2cm]{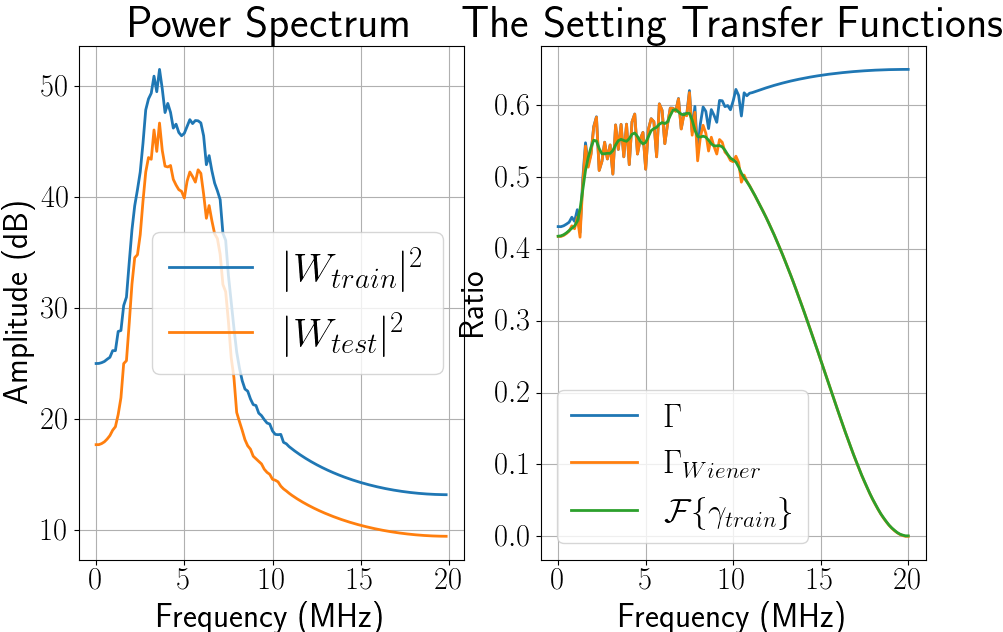}}\\
\end{tabular}
\caption{Calibration plots for focal location mismatches}
\label{fig:calib2}
\endgroup
\end{figure}

\begin{figure}[hbt!]
\begingroup
    \centering
    \begin{tabular}{c}
\hspace{-0.4cm}{ \includegraphics[width=8.8cm, height =4.2cm]{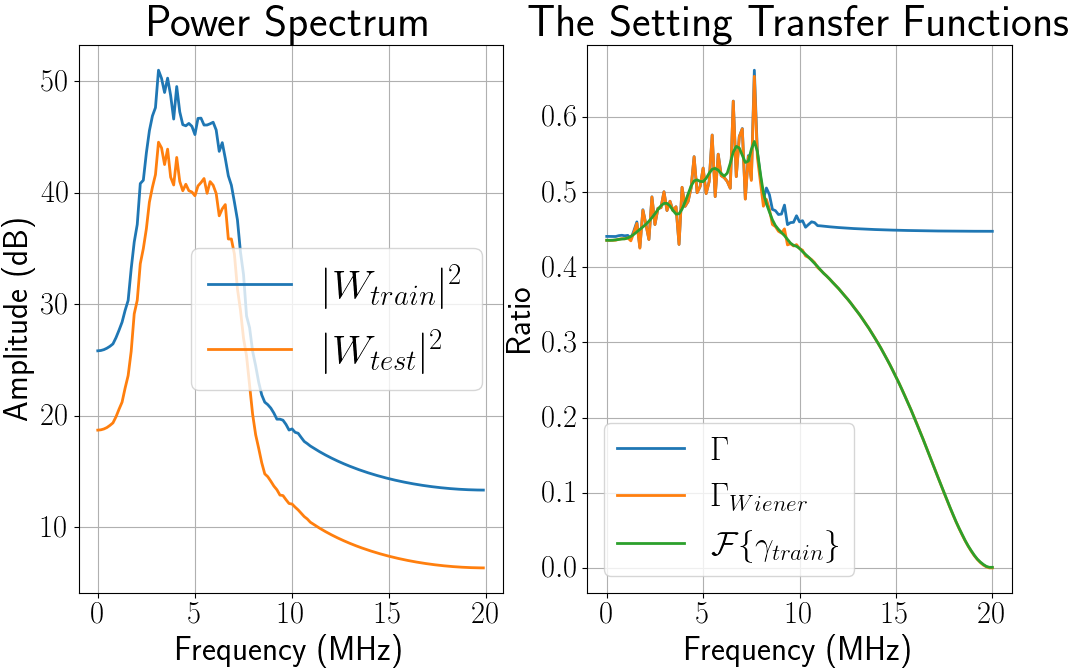}}\\
\end{tabular}
\caption{Calibration plots for the output power mismatches}
\label{fig:calib3}
\endgroup
\end{figure}

From Figs. \ref{fig:calib1}-\ref{fig:calib3}, we plotted power spectra for different training-testing setting pairs corresponding to Table \ref{tab:network} along with the setting transfer function $\Gamma(f,\textbf{x})$ and $\Gamma_{Wiener}(f,\textbf{x})$. Figure \ref{fig:calib1} is for the pulse frequency mismatch, Figs. \ref{fig:calib2} and \ref{fig:calib3} are for the focus mismatch and output power mismatch, respectively. These graphs are obtained from data acquired at a fixed axial location, which is around 2 cm. 
%In these figures, the top left sub-figures are the power spectra of the training and the testing settings obtained from Phantom1. The top right sub-figures are for power spectra of the training and the testing settings obtained from Phantom2. 
The left sub-figures show power spectra of the training and the testing settings obtained from the calibration. Furthermore, the right sub-figures depict $\Gamma(f,\textbf{x})$, $\Gamma_{Wiener}(f,\textbf{x})$ and Fourier Transform of the linear phase filter $\gamma_{train}$.

\vspace{-0.3cm}
\subsection{Transfer Learning}

Traditional transfer learning (TL) attempts to train and fine-tune a learning model when there is data mismatch\cite{sharif2014cnn, azizpour2015generic}. In this work, we fine-tuned a CNN, which had been trained using a large amount of data acquired from the samples at the training setting, by using a smaller amount of data acquired from the samples using the testing setting. Specifically, we compared the proposed method with TL for two different data set sizes (100 frames or 29 frames). For the first set, we fine-tuned the network using 50 frames for training, 25 frames for validation and 25 frames for testing. For the second, we used 15 frames for training, 7 frames in validation and 7 frames for testing. The major limitation for adopting the TL method, in comparison to the proposed method, is that it requires acquisition of a set of diverse frames from the actual samples at the the testing settings to be transferred to the model that was developed with data acquired using the training settings.

%% file: results.tex
\section{Results}
\label{sec:results}
\subsection{Pulse Frequency Mismatch}
\label{sec:results_pulse}
First, we investigated whether transfer functions could mitigate the effects of a frequency mismatch. We acquired the training data at a 9-MHz pulse frequency setting and the testing data at a 5-MHz pulse frequency setting.

\subsubsection{Train-Time Calibration} In Table \ref{tab:resa1}, \textit{“train-time calibration”} classification accuracies are listed when using different sets of data to train the model. For instance, row one indicates a probability of 0.5 meaning that 50\% of the data used in training of the model were from data acquired at the training alone and 50\% of the data used in the training of the model were from data collected at the training setting that were transferred to the testing setting. In the fourth row, a probability of 1 represents that 100\% of the data used in training the model were from data taken at the training setting that were transferred to the testing setting domain. Classification accuracy is given in mean classification accuracy $\pm$ standard deviation format. The mean classification accuracies ranged from 91.69\% to 95.30\%. For all experiments in Table \ref{tab:resa1}, we used 3e-7 as the learning rate. The epoch number was 25. 

\begin{table}[htb!]
    \centering
\caption{Train-Time Calibration for Pulse Frequency Mismatch \ref{sec:results_pulse}}
{\renewcommand{\arraystretch}{1.2}
{\small
\begin{tabular}{ |p{3.45cm}||p{2.2cm}|}
\hline
Experiment Type&  Accuracy \\
 \hline
Calibration w.p. 0.5& 91.69 $\pm$ 2.10\\
 \hline
Calibration w.p. 0.75& 91.70 $\pm$ 2.07\\
\hline
Calibration w.p. 0.9& 92.97 $\pm$ 1.33\\
 \hline
Calibration w.p. 1& 95.30 $\pm$ 1.40\\
\hline
\end{tabular}
}}
\label{tab:resa1}
\end{table}

\subsubsection{Test-Time Calibration vs Train-Time Calibration} In Table \ref{tab:resa2}, \textit{“train-time calibration”} and \textit{“test-time calibration”} for the pulse frequency mismatch are compared to a benchmark experiment in which there was no mismatch, i.e. the same pulse frequency was used for the training and testing data, to an experiment in which we didn't use any calibration. For \textit{train-time calibration}, we chose the experiment from Table \ref{tab:resa1} resulting in the highest mean classification accuracy, i.e., probability of 1. For \textit{test-time calibration}, we used 3e-7 as the learning rate with 25 epochs. For no calibration, we used 2e-8 as the learning rate with 25 epochs. In TL cases, we used 1.5e-7 as the learning rate with 30 epochs. Lastly, for the benchmark, we used 1.5e-7 as the learning rate with  30 epochs.

\begin{table}[htb!]
    \centering
\caption{Test-Time vs Train-Time for Pulse Frequency Mismatch \ref{sec:results_pulse}}
{\renewcommand{\arraystretch}{1.2}
{\small
\begin{tabular}{ |p{3.45cm}||p{2.2cm}|}
 \hline
Experiment Type&  Accuracy \\
 \hline
Train-Time Calibration& 95.30 $\pm$ 1.40\\
 \hline
Test-Time Calibration& 91.72 $\pm$ 1.54\\
\hline
TL with 50 Frames& 94.23 $\pm$ 1.66\\
\hline
TL with 15 Frames& 91.53 $\pm$ 3.24\\
\hline
No Calibration& 55.27 $\pm$ 3.83\\
\hline
Benchmark&99.61 $\pm$ 0.11\\
\hline
\end{tabular}
}}
\label{tab:resa2}
\end{table}

\subsection{Focus Mismatch}
\label{sec:results_focus}
In this subsection, we investigated whether transfer functions could mitigate effects of a focus mismatch. We acquired the training data focused at 2 cm and the test data with dual foci at 1 cm and 3 cm. 

Similar to the case with the pulse frequency mismatch, in train-time calibration, using only data transferred to the testing setting provided the highest classifier accuracy, i.e., 92.99\% for probability of 1 versus 86.94\% for probability of 0.5. In Table \ref{tab:resb2}, \textit{“train-time calibration”} and \textit{“test-time calibration”} for the focus mismatch are compared to the benchmark and no calibration experiments. For no calibration, we used 9e-8 as the learning rate with 25 epochs. In TL cases, we used 1.5e-7 as the learning rate with 30 epochs. For \textit{test-time calibration}, we used 2e-8 as the learning rate with 25 epochs. For \textit{train-time calibration}, we used 1e-8 as the learning rate with 25 epochs. The benchmark case was the same as the previous subsection.
\begin{table}[htb!]
    \centering
\caption{Test-Time vs Train-Time for Focal Mismatch \ref{sec:results_focus}}
{\renewcommand{\arraystretch}{1.2}
{\small
\begin{tabular}{ |p{3.45cm}||p{2.2cm}|}
 \hline
Experiment Type&  Accuracy\\
 \hline
Train-Time Calibration&  92.99 $\pm$ 1.21\\
 \hline
Test-Time Calibration& 88.57 $\pm$ 2.79\\
\hline
TL with 50 Frames& 89.18 $\pm$ 0.89\\
\hline
TL with 15 Frames& 76.06 $\pm$ 1.91\\
\hline
No Calibration& 64.36 $\pm$ 2.09\\
\hline
Benchmark& 99.61 $\pm$ 0.11\\
\hline
\end{tabular}
}}
\label{tab:resb2}
\end{table}

\subsection{Output Power Mismatch}
\label{sec:results_mix}

Finally, we investigated if we were able to mitigate effects of a data mismatch of output power by using transfer functions. We acquired the training data by using 0 dB output power, which represents the maximum output power of the imaging system, and the test data by using -6 dB output power, which represents the output power level that is 6 dB below the maximum. 

%\subsubsection{Train-Time Calibration}
%In Table \ref{tab:resc1}, we compare %\textit{“train-time calibration”} for the output %power mismatch. The mean classification accuracies %ranged from 90.71\% to 93.59\%. For all experiments %in Table \ref{tab:resb1}, we used 1e-7 as the %learning rate. The epoch number was 25.

%\begin{table}[htb!]
%    \centering
%\caption{Train-Time Calibration for Output Power Mismatch \ref{sec:results_mix}}
%{\renewcommand{\arraystretch}{1.2}
%{\small
%\begin{tabular}{ |p{3.45cm}||p{2.2cm}|}
% \hline
%Experiment Type&  Accuracy \\
% \hline
%Calibration w.p. 0.5&  90.71 $\pm$ 3.18  \\
 %\hline
%Calibration w.p. 0.75& 92.72 $\pm$ 2.04\\
%\hline
%Calibration w.p. 0.9& 92.90 $\pm$ 2.66\\
% \hline
%Calibration w.p. 1& 93.59 $\pm$ 1.87\\
%\hline
%\end{tabular}
%}}
%\label{tab:resc1}
%\end{table}

% \begin{table}[htb!]
%     \centering
% \caption{Label-Dependent vs Label-Free \ref{sec:results_mix}}
% {\renewcommand{\arraystretch}{1.4}
% {\small
% \begin{tabular}{ |p{3.45cm}||p{2.2cm}|}
%  \hline
% Experiment Type& Classification Accuracy\\
%  \hline
% Label-Dependent w.p. 0.5&  95.31 $\pm$ 1.16\\
%  \hline
% Label-Dependent w.p. 0.75& 93.84 $\pm$ 2.03\\
% \hline
% Label-Dependent w.p. 0.9&93.82 $\pm$ 1.70 \\
%  \hline
% Label-Dependent w.p. 1&91.64 $\pm$ 1.06\\
% \hline
% Label-Free w.p. 0.5& 94.45 $\pm$ 1.06 \\
%  \hline
% Label-Free w.p. 0.75& 94.20 $\pm$ 1.60\\
% \hline
% Label-Free w.p. 0.9& 92.48 $\pm$ 2.23 \\
%  \hline
% Label-Free w.p. 1&84.16 $\pm$ 1.98\\
% \hline
% \end{tabular}
% }}
% \label{tab:resc1}
% \end{table}

Similar to the case with the pulse frequency mismatch, in train-time calibration, using only data transferred to the testing setting provided the highest classifier accuracy, i.e., 99.32\% for probability of 1 versus 99.09\% for probability of 0.5. In Table \ref{tab:resc2}, \textit{“train-time calibration”} and \textit{“test-time calibration”} for the output power mismatch are compared to the benchmark and no calibration experiments. For no calibration, we used 5e-8 as the learning rate with 25 epochs. In TL cases, we used 1.5e-7 as the learning rate with 30 epochs. For \textit{test-time calibration}, we used 1.5e-7 as the learning rate with 30 epochs. For \textit{train-time calibration}, we used 1e-7 as the learning rate with 25 epochs. The benchmark was the same as the previous subsections.
% \begin{table}[htb!]
%     \centering
% \caption{Test-Time vs Training-Time \ref{sec:results_mix}}
% {\renewcommand{\arraystretch}{1.4}
% {\small
% \begin{tabular}{ |p{3.45cm}||p{2.2cm}|}
%  \hline
% Experiment Type& Classification Accuracy\\
%  \hline
% Train-Time Calibration& 95.31 $\pm$ 1.16 \\
%  \hline
% Test-Time Calibration& 81.76 $\pm$ 1.56\\
% \hline
% No Calibration&55.86 $\pm$ 4.51 \\
% \hline
% Benchmark& \\
% \hline
% \end{tabular}
% }}
% \label{tab:resc2}
% \end{table}

\begin{table}[htb!]
    \centering
\caption{Test-Time vs Train-Time for Output Power Mismatch \ref{sec:results_mix}}
{\renewcommand{\arraystretch}{1.2}
{\small
\begin{tabular}{ |p{3.45cm}||p{2.2cm}|}
 \hline
Experiment Type& Accuracy\\
 \hline
Training-Time Calibration& 99.32 $\pm$ 0.27  \\
 \hline
Test-Time Calibration& 99.50 $\pm$ 0.17 \\
\hline
TL with 50 Frames& 97.44 $\pm$ 0.70\\
\hline
TL with 15 Frames& 90.11 $\pm$ 2.29\\
\hline
No Calibration& 70.32 $\pm$ 6.10\\
\hline
Benchmark& 99.61 $\pm$ 0.11\\
\hline
\end{tabular}
}}
\label{tab:resc2}
\end{table}

%% file: discussions.tex
\section{Discussion}
\label{sec:discussion}

From Figs. \ref{fig:calib1}-\ref{fig:calib3}, the effect of the data mismatches on the averaged power spectrum are visualized. In Fig. \ref{fig:calib1}, the averaged power spectrum of the data from the training setting was shifted to higher frequencies in comparison to the averaged power spectrum of the data from the testing setting. That is expected because we used a 9-MHz pulse frequency in the training setting and a 5-MHz pulse frequency in the test setting. However, the actual shift in the spectrum was relatively narrower than 4 MHz. Overall, while the averaged power spectrum of the data from the testing setting was shifted to lower frequencies, the averaged power spectrum of the data from the training setting was shifted to a higher frequency. That led to the setting transfer functions $\Gamma$ and $\Gamma_{wiener}$ to be greater than unity below 5 MHz and less than unity above 5 MHz. $\Gamma_{wiener}$ matched $\Gamma$ well around the analysis bandwidth due to high SNR and rapidly goes to zero above 10 MHz.

In Fig. \ref{fig:calib2}, the averaged power spectrum of the data from training setting had higher amplitude than the averaged power spectrum of the data from testing setting because those plots were obtained around 2 cm axially, which corresponds to the focal region of the training setting. As a result, the setting transfer functions $\Gamma$ and $\Gamma_{wiener}$ were approximately constant at 0.6 around the analysis bandwidth. Similarly, in Fig. \ref{fig:calib3}, the averaged power spectrum of the data from the training setting had higher amplitude than the averaged power spectrum of the data from the testing setting. That is due to using 6 dB higher output power in the data acquisition, which led to the setting transfer functions $\Gamma$ and $\Gamma_{wiener}$ to be relatively constant at 0.5 around the analysis bandwidth. Similar to Fig. \ref{fig:calib1}, in Fig. \ref{fig:calib2} and Fig. \ref{fig:calib3}, $\Gamma_{wiener}$ approached zero above 10 MHz due to low SNR. 

In Table \ref{tab:resa1}, we observed that as the probability level increased, the mean classification accuracy increased for the given mismatch. The probability levels indicated how much of the data was calibrated in the training. For instance, when we applied \textit{train-time calibration} for half of the data (calibration w.p. 0.5), i.e. half of the training data was calibrated, the mean classification accuracy became 91.69\%. When we applied \textit{train-time calibration} for all the training data (calibration w.p. 1), the mean classification accuracy became 95.30\%. This indicates that calibrated training data provided all the potential performance increase. Using mismatched (uncalibrated) training data did not provide any additional performance increases for the given experiments. 

In Table \ref{tab:resa2}, we observed that the proposed method mitigated the effects of the given frequency mismatch. The CNN trained without any calibration resulted in a mean classification accuracy of 55.27\%, which is not much better than a random guess classifier. Further, we obtained mean classification accuracies of 95.30\% and 91.72\% for \textit{train-time calibration} and \textit{test-time calibration}, respectively, which are substantially closer to the benchmark performance. Furthermore, TL couldn't catch up to the proposed calibration method, which used 10 fixed frames, even with 50 diverse frames. Hence, TL was more expensive and a less accurate approach than the proposed calibration method for the given task.

Lastly, \textit{train-time calibration} performed better than \textit{test-time calibration} for the given mismatch in two of the three cases. A way to interpret it could be that augmenting the training data led to a decrease in data mismatch in the training. This could provide some algorithmic advantages, such as choosing hyper-parameters.  

In Table \ref{tab:resb2}, we observed that the proposed method mostly mitigated the effects of the given focus mismatch. While the CNN trained without any calibration resulted in a mean classification accuracy of 64.36\%, which is slightly better than a random guess classifier. We obtained mean classification accuracies of 92.99\% and 88.57\% for \textit{train-time calibration} and \textit{test-time calibration}, respectively, which were substantially closer to the benchmark performance. Lastly, \textit{train-time calibration}, which used only 10 fixed frames, performed better than transfer learning with 50 diverse frames.

In Table \ref{tab:resc2}, we observed that the proposed method mitigated the effect of the given output power mismatch. While the CNN trained without any calibration resulted in a mean classification accuracy of 70.32\%, we obtained mean classification accuracies of 99.32\% and 99.50\% for \textit{train-time calibration} and \textit{test-time calibration}, respectively, which are at the benchmark performance. Unlike for results from the previous setting mismatches, \textit{test-time calibration} performed slightly better than \textit{train-time calibration}. Lastly, \textit{train-time calibration}, which used only 10 fixed frames, performed better than transfer learning with 50 diverse frames.

One advantage of \textit{test-time calibration} over \textit{train-time calibration} is its simplicity. The \textit{train-time calibration} requires the training data to be converted, added to the data, and the model retrained. With the \textit{test-time calibration}, there is no need for any retraining or fine tuning of the network when a new test setting is being used. Therefore, the \textit{test-time calibration} approach is more suitable for real-time clinical applications. 

%Overall, \textit{label-dependent calibration} performed better than \textit{label-free calibration} for the given frequency and output power mismatches while it under-performed \textit{label-free calibration} for the given focus mismatch. If we assume that the imaging process is linear with similar noise characteristics through out the experiments, we wouldn't expect to identify any difference between \textit{label-dependent calibration} and \textit{label-free calibration}. However, as we observed from the results, \textit{label-dependent calibration} had slight advantage for the given mismatches. That can be attributed to \textit{label-dependent calibration} being more tolerant to non-linearities and deviations in noise characteristics. 
Overall, as a higher percentage of calibrated data were used in the \textit{train-time calibration} approach, the classifier performance increased. When the classifier used 100\% calibrated data, it achieved its best performance, which indicates that using mismatched or uncalibrated training data doesn't provide any additional performance increase for the given mismatches. However, in the general case, we could consider using uncalibrated data in training by adjusting the ratios of calibrated and uncalibrated data. Depending on the type of applications and data mismatches, using uncalibrated data in training could potentially provide richer training data and better generalizability. 

Regarding \textit{test-time calibration} vs \textit{train-time calibration}, we observe that \textit{train-time calibration} performed better for more complex mismatches such as the frequency mismatch and the focus mismatch. However, \textit{test-time calibration} performed better for the output power mismatch, which could be calibrated by using a simple scale factor. Hence, this limited study suggests that \textit{train-time calibration} is preferable when the data mismatch is complex. 

%Another aspect of \textit{train-time calibration} is that it has algorithmic advantages over \textit{test-time calibration} because it has more relevant training error, which makes hyper-parameter selection easier. On the other hand, \textit{test-time calibration} requires no retraining or no fine tuning of the model for each test setting, and therefore, it is more desirable for real-time applications than \textit{train-time calibration}. 

%One observation from the right bottom sub-figure of Fig. \ref{fig:calib1} is that the setting transfer function $\Gamma$s obtained from Phantom1, Phantom2 and the calibration phantom, followed very similar patterns. 
One important future work would be related to how to select the calibration phantom and its properties. When non-linearities in the system and imaging substrate were negligible, the system response of an ultrasound system and hence the setting transfer function, $\Gamma$, should be the same irrespective of the imaging substrate. Therefore, we could use any phantom with uniform scattering properties as the calibration phantom. However, due to random spatial variation noise from the subresolution scatterers, there could be some fluctuation in the setting transfer function $\Gamma$ as it can be observed from Figs. \ref{fig:calib1}-\ref{fig:calib3}. If one could average multiple views, variation in the power spectra would decline. However, it would also increase the complexity of the approach. In the proposed approach, we acquired multiple frames of the same view after stabilizing the transducer for the calibration data. This way, the acquisition of the calibration data could be automated easily for all imaging settings without any human intervention. The need of multiple views in the calibration data would make it more complicated and expensive.  
%The sub-figure supports the hypothesis that the choice of phantom did not matter for the case of the frequency mismatch. 

Lastly, among the investigated acquisition-related data mismatches, calibrating the focus mismatch led to the lowest mean classification accuracy, which was 92.99\%. The frequency mismatch led to largest drop in classification accuracy for the no calibration case. The mean classification accuracy dropped to 55.27\% from 99.61\%. These observations demonstrate that a frequency mismatch and a focus mismatch are indeed more challenging than an output power power mismatch.

%% file: conc.tex
\section{Conclusion}
\label{sec:conclusion}
We demonstrated that the proposed approach for mitigating the effects of data mismatches was effective for tissue classification  under various mismatches in training versus testing scanner settings: a pulse frequency mismatch, a focal region mismatch and an output power mismatch. Therefore, the incorporation of transfer functions between scanner settings can provide an economical means of generalizing a DL model for the specific imaging tasks where scanner settings are not fixed by the operator.